\documentclass[sn-mathphys-num]{sn-jnl} 

\usepackage{graphicx}%
\usepackage{multirow}%
\usepackage{amsmath,amssymb,amsfonts}%
\usepackage{amsthm}%
\usepackage{mathrsfs}%
\usepackage[title]{appendix}%
\usepackage{xcolor,colortbl}%
\usepackage{textcomp}%
\usepackage{manyfoot}%
\usepackage{booktabs}%
\usepackage{algorithm}%
\usepackage{algorithmicx}%
\usepackage{algpseudocode}%
\usepackage{listings}%

\definecolor{Gray}{gray}{0.85}
\definecolor{LightCyan}{rgb}{0.88,1,1}

\newcolumntype{a}{>{\columncolor{Gray}}c}
\newcolumntype{b}{>{\columncolor{white}}c}
\newcolumntype{d}{>{\columncolor{Gray}}l}

\newcommand\T{\rule{0pt}{2.6ex}}       
\newcommand\B{\rule[-1.2ex]{0pt}{0pt}} 

\begin{document}

\title{Optimizing Qubit Control Pulses for State Preparation
}

\author[1]{\fnm{Annika S.} \sur{Wiening}}
\author[1]{\fnm{Jörn} \sur{Bergendahl}}
\author*[2]{\fnm{Vicente} \sur{Leyton-Ortega}}\email{leytonorteva@ornl.gov}
\author*[1]{\fnm{Peter} \sur{Nalbach}}\email{Peter.Nalbach@w-hs.de}

\affil[1]{
\orgdiv{Fachbereich Wirtschaft \& Informationstechnik}, 
\orgname{Westf\"alische Hochschule}, \orgaddress{M\"unsterstrasse 265}, 
\postcode{46397},
\city{Bocholt},
\state{North Rhine-Westphalia}, 
\country{Germany}
}

\affil[2]{
\orgdiv{Quantum Computational Science Group, Quantum Information Science Section, Computational Sciences and Engineering Division}, 
\orgname{Oak Ridge National Laboratory}, 
\city{Oak Ridge}, 
\state{Tennessee}, 
\postcode{37831}, 
\country{USA}
}

\abstract{
In the burgeoning field of quantum computing, the precise design and optimization of quantum pulses are essential for enhancing qubit operation fidelity. This study focuses on refining the pulse engineering techniques for superconducting qubits, employing a detailed analysis of {\it Square} and {\it Gaussian} pulse envelopes under various approximation schemes. We evaluated the effects of coherent errors induced by naive pulse designs. We identified the sources of these errors in the Hamiltonian model's approximation level. We mitigated these errors through adjustments to the external driving frequency and pulse durations, thus, implementing a pulse scheme with stroboscopic error reduction. Our results demonstrate that these refined pulse strategies improve performance and reduce coherent errors. Moreover, the techniques developed herein are applicable across different quantum architectures, such as ion-trap, atomic, and photonic systems.
}

\keywords{Quantum Gate Pulse Design, Quantum Computing, Chirp Modulation, Superconducting Qubits, Coherent Errors}

\maketitle

\date{}

\section{Introduction}

Recently, we have seen big steps forward in quantum computing: increased qubit count and coherence time. Among these are IBM’s Eagle, Heron, and Osprey quantum chips with 127, 133, and 433 qubits \cite{chow2021ibm}; IQM’s processors from 20 to 54 qubits \cite{IQM}; Rigetti’s ANKAA and Aspen chips with 80 or more qubits \cite{Rigetti}; and a newly proposed 32-qubit ion-trap-based quantum computer from Quantinuum \cite{moses2023race}, among the others \cite{gill2024quantum}. Moreover, ongoing ambitious efforts are to introduce quantum computing systems based on different technologies \cite{singh2024survey} such as photonics (Xanadu \cite{Xanadu}, PsiQuantum \cite{psiquantum}, and Quandela \cite{quandela}), neutral atoms (PASQAL \cite{pasqal} and Atom Computing \cite{atomcomputing}), and quasi-particles (Microsoft \cite{gibney2016inside}, Google \cite{quantumai}, and Quantinuum \cite{iqbal2024non}). These factors increase quantum volume, making the quantum circuits more complex and large. The quantum computing industry is on the brink of an era when large circuits with many layers are executed, and many gates will soon become a possibility in the NISQ-like regime. This advance enables new applications but also creates new challenges, increasing the importance of pulse design for quantum gates. In particular, high-precision pulse engineering gains special importance as quantum circuits become more complex.

The design of pulses is essential to the success of quantum computing, placing a burden on every aspect of quantum, from qubit initialization to performing complex algorithms. It generates complex choreographies, coping with the effect upon quantum states of fine-tuned energy deposition. Every pulse must be crafted to enhance qubit coherence and suppress errors while also dealing with the inherent properties of the qubits and their interactions with the environment. By dealing with larger quantum systems, getting hands-on pulse design at a more profound level is helpful and necessary for the accuracy of quantum computers in executing complex quantum algorithms. There has been a dramatic growth in pulse design, which nowadays takes advantage of advanced quantum control techniques - e.g., GRAPE (Gradient Ascent Pulse Engineering) \cite{khaneja2005optimal} and GOAT (Gradient Optimization of Analytic conTrols) \cite{machnes2018tunable}. Most of these methods require extensive hardware characterization or demanding simulations that can model quantum operations close to the error limits.

Researchers have been investigating many pulse-shaping methods to address these needs to reduce errors and increase gate fidelities, especially for superconducting qubits. As an example, Slepian pulses (also known as DRAG -derivative removal by the adiabatic gate- pulses) \cite{gambetta2011analytic} are very common experimentally since their sharp spectral profile provides a method to limit excitation of the qubit to higher energy states that won't be used in the computation. Extending this theme, pseudo-Chebyshev \cite{dorozhovets2023weight} pulse candidates have been introduced for two-qubit gates, which offer significantly improved gate fidelities compared to Slepian pulses.

A multiplexed control architecture \cite{zhao2024multiplexed} has been suggested to overcome the wiring challenges associated with increasing the number of qubits. Each row-column intersection commands a unique set of pulses, while the addressing and controlling are parallel for each qubit with the shared row-column control lines used in this system. This exceedingly simple dynamic protocol enables the removal of intricate wiring complexities such as those encountered in large-scale quantum processors.

Additionally, improvements in pulse generator architecture, such as SPulseGen \cite{matsuo2023spulsegen}, offer substantial simplification in the quantum control hardware landscape. This allows for savings in cost and complexity of the control electronics since the arbitrary waveforms generator (AWGs), which are expensive, can be avoided using simple square-shaped pulses. Pulses from nuclear magnetic resonance, including Composite \cite{brown2004arbitrarily,alway2007arbitrary} and Adiabatic pulses \cite{saffman2020symmetric,williams2024quantification} — hyperbolic secant and B1-insensitive rotation pulses, for example \cite{williams2024quantification} — have also been retooled to increase the accuracy and speed of quantum computing operations\cite{jones2024controlling,williams2024quantification}.

Building on these advances, this work specifically focuses on the pulse design for superconducting qubits by fine-tuning the external control signals in frequency and duration. By comparing different optimization approaches for the Square and Gaussian pulse profiles, we determine at first the coherent errors due to the baseline rotating wave approximation. We then extend our analysis to include the first level corrections of this corresponding Magnus expansion and, thereby, reducing the coherent errors by several orders of magnitude for nearly identical pulse durations and powers.

In particular, we consider the design of $Y_{\pi/2}$ and $Y_{\pi}$ pulses, adopting the notation $A_\theta = \exp [-i \sigma_A \theta/2]$, where $\sigma_A$ represents the Pauli matrices $\sigma_x, \sigma_y$ and $\sigma_z$. These quantum operations are essential for preparing one-qubit quantum states and basic quantum operations (local gates), as the general rotation $U3(\phi, \theta, \lambda)$ can be decomposed into $Z_{\phi - \pi/2} X_{\pi/2} Z_{\pi - \theta} X_{\pi / 2} Z_{\lambda - \pi/2}$ \cite{mckay2017efficient}. Here, $Z$-rotations as virtual operations, effectuate phase shifts in the local oscillator phase. The $X$ rotations are realized using $X_\phi = Z_{-\pi/2} Y_\phi Z_{\pi/2}$. 
This underlines the importance of precise $Y_{\pi/2}$ pulse design in performing a general local rotation $U3$. 

In Section \ref{sec:quantumdynamics}, we introduce the Hamiltonian model, defining the dynamics and control mechanisms for superconducting qubits and detailing the various pulse types such as square 
and Gaussian pulses. In Section \ref{Ypi}, we explore the configuration and implementation of $Y_{\pi}$ pulses, examining the impacts of both the rotating wave approximation and the first order corrections of the Magnus expansion, with a particular focus on the coherent errors introduced by these approximation schemes. Section \ref{xpihalf} is dedicated to the  $Y_{\pi/2}$ pulse design, highlighting their role in achieving high-fidelity quantum operations. Moving forward, Section \ref{sec:stateprep} discusses strategies for initial state preparation, illustrating how precise pulse engineering enhances the accuracy and reliability of quantum state initialization. We conclude in Section \ref{sec:conclusions}, summarizing our findings and discussing their implications for future research in quantum computing, particularly in the context of scaling up quantum operations.

\section{Quantum Dynamics and Control in Superconducting Qubits} \label{sec:quantumdynamics}

Typically, superconducting qubits are modeled as Duffing oscillators, including beyond the ground $|0 \rangle$ and the excited state $|1 \rangle$ further higher energy states. We, however, restrict our consideration to the quantum two-level regime for each qubit. For a single qubit, we, thus, obtain the Hamiltonian ($\hbar=1$)
\begin{equation}
    H = -\frac{\omega_{q}}{2} \sigma_{z}+\Omega_{d}D(t)\sigma_{x} \ , 
\end{equation}
with the qubits resonance frequency $\omega_q$ and the driving energy scale $\Omega_d$ (physically determined by applied field strength times addressed moment). 
The qubit resonance frequency is cavity-dressed but not necessarily what is returned, for example, by the backend (quantum device) defaults of IBM machines. The latter includes the dressing due to the qubit-qubit interactions on a multi-qubit chip. Quantities are given in angular frequencies, with units $2\pi$ GHz. 
Qubits are addressed via drive channels with a signal (drive amplitude) $d(t)$ imposed on an oscillating carrier, i.e.
\begin{equation}
    D(t) = d(t) \sin(\omega_{\rm LO} t + \phi_{\rm LO}) \quad\mbox{with}\quad |D(t)|\le 1 .
\end{equation}

As typical parameters, we employ values from the \texttt{IBMQ-Manila} backend, i.e., $\omega_q\simeq 30\, {\rm G Hz}$ which reflect roughly qubits with frequency $f = \omega_q / (2 \pi) \simeq 5\, {\rm GHz}$. Typical couplings to neighbors are $J\simeq 10\, {\rm MHz}$ and driving energies $\Omega_d\simeq 1\, {\rm GHz}$ which corresponds to weakly driven qubits, i.e. $\Omega_d \simeq 3 \times 10^{-2} \omega_q$. \footnote{The \texttt{IBMQ-Manila} backend configuration provides the following information for qubit 0: $\omega_{q,0} =29806862687.393623\, {\rm Hz}$, $\Omega_{d,0} =  982583670.175613\, {\rm Hz}$, and $J_{0,1} = 13906241.266973624 \, {\rm Hz}$ for the coupling to its neighbor.}
The used cavities restrict the available driving frequencies to the vicinity of the resonance with $\Delta\omega_{\rm LO} \simeq 2\pi\cdot 1\, {\rm GHz}$.

For weak and near-resonant driving, $\Omega_d/\omega_{\rm LO}\ll 1$, and detuning $\delta = \omega_q - \omega_{\rm LO}$ which is similarly small, i.e. $\delta/\omega_{\rm LO}\ll 1$, the dynamics of the driven qubit can be efficiently described within the rotating frame \cite{thimmel1999rotating}. Assuming that the pulse amplitude $d(t)$ varies slowly over time, such that $|d(t+\omega_{\rm LO}^{-1}) - d(t)|\ll (\Omega_d/\omega_{\rm LO})$, traditional analysis employs the rotating wave approximation (RWA) to simplify the Hamiltonian. However, to account for higher-order effects not captured by the traditional RWA, we introduce an enhanced version, termed RWA+ \cite{thimmel1999rotating}. This extended approximation incorporates first-order corrections in the small parameters, leading to a more accurate description of the system dynamics:
\begin{equation}\label{Heff}
H_{\rm eff} = \left[ -\delta - \frac{3 (\Omega_d d(t))^2}{4 \omega_{\rm LO}}\right] \frac{\sigma_x}{2}
+ \Omega_d  d(t) \left[ 1 - \frac{\delta}{2\omega_{\rm LO}}\right] \frac{\sigma_y}{2} .
\end{equation}
In the lowest order, i.e., the traditional RWA, and particularly at resonance ($\delta=0$), the effective Hamiltonian simplifies to
\begin{equation}\label{eq:rwa}
H^0_{\rm eff} = \Omega_d  d(t)  \frac{\sigma_y}{2} ,
\end{equation}
which describes a Rabi rotation with the Rabi frequency $\Omega_d d(t)$ around the $y$-axis of the Bloch sphere. The modifications in Eq. (\ref{Heff}) under RWA+ refine the description by adjusting the resonance condition and altering the Rabi frequency, thus enhancing the fidelity of qubit control.

The resonance is shifted by the Bloch-Siegert shift $ 3(\Omega_d d(t))^2 / (4\omega_{\rm LO})$ resulting in a slightly blue-shifted resonance frequency
\begin{equation}
    \frac{\omega_{\rm LO, res}}{\omega_q}  \simeq   1 + \frac{3(\Omega_d \, d(t))^2}{4\, \omega_q^2} +O\left( \left(\frac{\Omega_d}{\omega_q} \right)^4 \right) \ ,
\end{equation}
where $\omega_{\rm LO, res}$ is the shifted resonance frequency. 

The Rabi frequency is adjusted to account for the shifted resonance:
\begin{equation}
    \Omega_d \, d(t) \left[ 1 - \frac{\delta}{2\, \omega_{\rm LO}}\right]
    \simeq \Omega_d\, d(t) \left[ 1 + \frac{3(\Omega_d \, d(t))^2}{8\, \omega_q^2} +O\left( \left(\frac{\Omega_d}{\omega_q} \right)^4 \right)\right] ,
\end{equation}
This formula indicates that the Rabi frequency increases slightly when the system is driven at the Bloch-Siegert shifted resonance.

%
%

%
%

\subsection{Driving Envelope Configurations}

Having computed the first order corrections in the Magnus expansion for a given driving function $d(t)$, we now explore three distinct types of driving configurations. As our initial configuration, we consider the \textit{Square Envelope} defined in terms of a Heavyside function $\Theta(t)$ as:
\begin{equation}
    d_{\rm sq}(t) = \Theta(t)\Theta(T-t) \ ,
\end{equation}
where $t=0$ marks the beginning and $t=T$ the end of the pulse with duration $T$. Although the square pulse is conceptually simple and often used for theoretical analysis, it is generally avoided in practical applications due to its tendency to cause sharp transitions that can lead to population losses into higher excited states.

As the second envelope configuration, we will consider the common \textit{Gaussian Envelope} expressed as:
\begin{equation}
    d_{\rm ga}(t) = e^{-\frac{1}{2} \frac{(t-T/2)^2}{\sigma^2}} \Theta(t)\Theta(T-t) \ ,
\end{equation}
where $\sigma$ represents the Gaussian width. This envelope starts and ends with non-zero but small amplitudes at $t=0$ and $t=T$, respectively, making it smoother and more experimentally favorable than the {\it Square Envelope}.

To achieve zero amplitude at the start and end of the pulse, we consider the \textit{Shifted Gaussian Envelope}, a Gaussian envelope modified as follows:
\begin{equation}
    d_{\rm sga}(t) = \frac{e^{-\frac{1}{2} \frac{(t-T/2)^2}{\sigma^2}} - d_{\rm ga}(0)}{1- d_{\rm ga}(0)} \Theta(t)\Theta(T-t)
\end{equation}
where $d_{\text{ga}}(0) = e^{-\frac{1}{2} \frac{(T/2)^2}{\sigma^2}}$. This shifted Gaussian ensures the pulse amplitude smoothly transitions from and back to zero, minimizing abrupt changes.

Additionally, Derivative Removal by Adiabatic Gate (DRAG) pulses offers further improvements. These utilize multi-Gaussian profiles to drive qubit transitions via both $\sigma_x$ and $\sigma_y$ directions, reducing gate errors significantly. 

Despite limiting our discussion to the two-level system approximation, where population losses to higher states are theoretically precluded, the choice of driving envelope significantly impacts gate fidelity. Our analysis covers square, Gaussian, and shifted Gaussian pulses, with implications extendable to DRAG pulses due to their multi-Gaussian nature. This study illustrates how different pulse shapes influence the effectiveness and errors of quantum gates even without considering population losses to higher excited states.

\section{$Y_{\pi}$ Gate Implementation} \label{Ypi}

Given the formulation of the effective Hamiltonian, we start implementing the $Y_{\pi}$ gate, a fundamental operation characterized by a $\pi$ pulse along the $y$-axis of the Bloch sphere. We will focus on two pulse envelope shapes, {\it Square} and {\it Gaussian}, under the RWA and its enhanced version RWA+.

\subsection{Square Envelope under RWA and RWA+}
Starting with the simplest scenario in the lowest order and at resonance $\delta=0$, where the dynamics is governed by the Hamiltonian \eqref{eq:rwa} with the evolution operator
\begin{equation}
    U(T_0) = \exp \left[-i \frac{\sigma_y}{2}\int_0^{T_0}  dt \, \Omega_d \, d(t) \right] .
\end{equation}
In this approximation, the qubit undergoes a flip when the accumulated phase equals $\pi$, defining a $\pi$-pulse. Thus, the period required to achieve this operation is $T_0 = \pi/\Omega_d$. 

In the enhanced approximation (RWA+), the pulse duration $T_1$ to achieve a qubit flip is modified: 
\begin{equation}
    T_1\cdot \Omega_d \left[ 1 - \frac{\omega_q - \omega_{\rm LO, res}}{2\omega_{\rm LO, res}}\right] = \pi \,\Leftrightarrow\, T_1 = \frac{\pi}{\Omega_d} \cdot \left( 1 - \frac{3\Omega_d^2}{8\omega_q^2} 
    +O\left( \left(\frac{\Omega_d}{\omega_q} \right)^4 \right)\right) .
\end{equation}
%



The Magnus expansion \cite{thimmel1999rotating} reveals that the effective Hamiltonian \eqref{Heff} describes the dynamics stroboscopically, i.e. only at 
 at times that are multiples of the inverse driving frequency the effective Hamiltonian correctly describes the dynamics. In the intervals between these times, additional corrections become significant which we avoid by a careful adjustment of the drive parameters. By fine-tuning the drive amplitude, we ensure that the pulse duration aligns with a multiple of the inverse driving frequency, thus enhancing the accuracy of our quantum control.

A substantial reduction in coherent error is observed only when both the Bloch-Siegert shift in the resonance frequency and the pulse duration, adjusted to a multiple of the inverse resonance frequency, are simultaneously considered. Employing only one of these corrections results in an overall coherent error comparable to the error observed with the rotating wave approximation alone.

\subsection{Gaussian Envelope under RWA and RWA+}
Initially, we analyze the Gaussian envelope under the rotating wave approximation (RWA). For a Gaussian pulse, the integral of the pulse shape over time must satisfy the condition for a $\pi$ pulse:
\begin{equation}\label{condpi}
   \pi = \Omega_d \int_0^T dt e^{-\frac{1}{2} \frac{(t-T/2)^2}{\sigma^2}} \le \Omega_d \sigma \sqrt{2\pi} .
\end{equation}
For $T\gg \sigma$, the integral approaches the upper bound value. Hence, when setting a finite duration $T$, the width must be determined numerically. Alternatively, we fix a width and then adjust the duration. The same procedure must be used for the {\it Shifted Gaussian Pulse}. In detail, we fix the width to 1\% larger than the theoretical width for infinite-duration pulses. This leads to a duration roughly five times the width.

In technical realizations, a square pulse results in substantial leakage errors. The first improvement is the application of Gaussian pulses. The Bloch-Siegert shift for a Gaussian pulse profile (as well as for all more advanced schemes) is time-dependent. Thus, implementing corrections to the rotating wave approximation forces us to chirp the qubit, i.e., to drive the qubit with a changing frequency. Note that a rotating wave approximation scheme can also be determined for chirped systems \cite{nalbach2018magnus}.

To fix the pulse duration and width of the {\it Gaussian Envelope}, we now have to solve
\begin{equation}
    \int_0^T dt \Omega_d d(t) \left( 1 + \frac{3\Omega_d^2 d^2(t)}{8\omega^2_q} \right) = \pi .
\end{equation}
As before, we fix the duration $T$ and determine the width numerically. 

Since a time-dependent frequency does not allow to readily define a period, we cannot adjust the duration to a multiple of a drive period. Instead we ensured that the $\sin$-wave passes through a zero (from negative to positive) at the end of the duration. These corrections did not result in any suppression of the error.

We then determined the average resonance frequency during the pulse resulting in
\begin{equation}
    \bar{\omega}_{\rm LO, res} \simeq  \omega_q \left( 1 + \frac{3\Omega^2_d }{4\omega_q^2} c_1 \right) \quad\mbox{with}\quad c_1 = \frac{1}{T}\int_0^T dt d^2(t) \simeq \sqrt{\pi} \frac{\sigma}{T}
\end{equation}
for $\sigma\ll T$. In our simulation, typically, $5\sigma\lesssim T\lesssim  6\sigma$ and, thus, $0.295\lesssim c_1\lesssim 0.355 $. Using this time-independent frequency results in a pulse definition
\begin{equation}
   \left( 1 + \frac{3\Omega_d^2 }{8\omega^2_q} c_1 \right)  \int_0^T dt \Omega_d d(t) = \pi \;\Leftrightarrow\; \int_0^T dt \Omega_d d(t) = \pi \left( 1 - \frac{3\Omega_d^2 }{8\omega^2_q} c_1 \right) 
   +O\left( \left(\frac{\Omega_d}{\omega_q} \right)^4 \right)
\end{equation}

\subsection{Coherent errors}

\begin{figure}
\centering
\includegraphics[width=8.5cm]{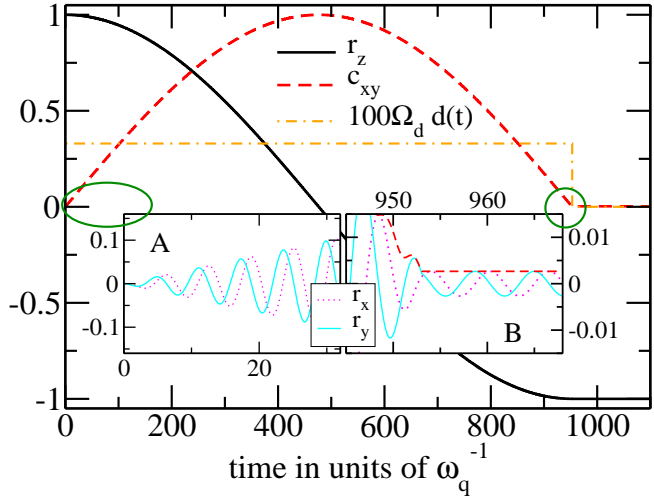}
\caption{\label{fig1} Numerical experiment results of a $\pi$-pulse using a square pulse profile, showing population inversion from $r_z=1$ to $r_z=-1$ (depicted by the black full line in the main figure). The coherence amplitude $c_{xy}=(r_x^2 + r_y^2)^{1/2}$, represented by the red dashed line, starts at zero and ideally should return to zero at the pulse’s conclusion at time $953 \omega_q^{-1}$. Despite this expectation, finite coherence at the pulse end highlights a coherent error, as detailed in inset B. Oscillations of $r_x$ and $r_y$ with frequency $\omega_q$ are shown in inset A. The amplitude of the square pulse, scaled up by a factor of 100 for visibility, is shown in units of $\omega_q$. Time is measured in units of $\tau_q=\omega_q^{-1}$, providing a scale for the temporal dynamics involved.}
\end{figure}

Implementing a $\pi$-pulse introduces finite coherent errors due to the necessity of switching to the rotating frame for harmonic driving. This approach is viable in a two-level system, where there is no concern about leakage into higher energy states, allowing the use of square pulses without needing to minimize gradients responsible for leakage errors. Employing qubit parameters from the \texttt{IBMQ-Manila} backend, we calculated the residual coherence, $c_{xy}=\sqrt{r_x^2+r_y^2}$ (with expectation values $r_i=\langle \sigma_i\rangle$), as detailed in Table \ref{table:Ypi}. Fig. \ref{fig1} shows exemplary a square $\pi$ - pulse (dash-dotted orange line) and the profile for the coherence $c_{xy}$ (dashed red line) and population $r_z$ (full black line). The two insets show the start and the end of the pulse in detail. The latter shows a finite final coherence (which reflects the coherent error) for this RWA pulse.

Table \ref{table:Ypi} evaluates coherent errors across various driving amplitude levels ($\Omega_d$) under different optimization scenarios, showcasing how each approach impacts error reduction. These scenarios include the basic Rotating Wave Approximation (RWA), which serves as the minimal correction baseline; adjustments for full periods within the RWA to align pulse duration with the driving frequency; the addition of higher-order corrections (RWA + Corrections) to refine the Hamiltonian; and sophisticated corrections that combine higher-order terms with period adjustments (RWA + Corrections, Full Periods). We also applied the most comprehensive approach, RWA + Effective Corrections, Full Periods, which utilizes a numerically optimized effective frequency shift (smaller than the Bloch-Siegert shift) for maximum error reduction, significantly enhancing control precision, especially at lower driving amplitudes.

\begin{table}[t]
    \caption{$Y_\pi$-pulse Coherent error $c_{xy}$: \label{table:Ypi}}
    \centering
    \begin{tabular}{|d|b|b|b|} \hline
    \rowcolor{LightCyan}
       Square $\pi$ - Pulse & $\Omega_d=0.2\Omega_d^{(max)}$ & $\Omega_d=0.1\Omega_d^{(max)}$ & $\Omega_d=0.05\Omega_d^{(max)}$ \T\B \\ \hline
       RWA  &  $5.9\cdot 10^{-3}$ &  $2.7\cdot 10^{-3}$ & $1.6\cdot 10^{-3}$ \T\B \\ \hline
       RWA, full periods &  $1\cdot 10^{-2}$ &  $4.9\cdot 10^{-3}$ & $2.5\cdot 10^{-3}$ \T\B \\ \hline
       RWA + corrections &  $5.6\cdot 10^{-3}$ &  $3\cdot 10^{-3}$ & $1.3\cdot 10^{-3}$ \T\B \\ \hline
       RWA + corrections, full periods &  $1.4\cdot 10^{-4}$ &  $4.2\cdot 10^{-5}$ & $1.4\cdot 10^{-5}$ \T\B \\ \hline
       RWA + eff. corrections, full periods &  $1.4\cdot 10^{-4}$ &  $3.6\cdot 10^{-5}$ & $9\cdot 10^{-6}$ \T\B \\ \hline
       \rowcolor{LightCyan}
        Gaussian $\pi$ - Pulse &  & & \T\B \\ \hline
       \begin{tabular}{l} RWA \\ RWA, full periods \\ 
                          RWA + $t$-dep. corrections + zero cross
        \end{tabular}   &  $2.8\cdot 10^{-3}$ & $1.4\cdot 10^{-3}$ & $7\cdot 10^{-4}$ \T\B \\ \hline
       RWA + eff. mean corrections &  $10^{-3}$ & $5\cdot 10^{-4}$ & $2.5\cdot 10^{-4}$ \T\B \\ \hline
       RWA + eff. opt. corrections &  $1.6\cdot 10^{-4}$ & $ 10^{-5}$ & $5\cdot 10^{-5}$ \T\B \\ \hline
       RWA + eff. opt. corrections, full periods &  $1.6\cdot 10^{-5}$ & $4\cdot 10^{-6}$ & $10^{-6}$ \T\B \\ \hline
    \end{tabular}
    Quantitative evaluation of coherent errors $c_{xy}$ for a Square and Gaussian $\pi$-pulse under different optimization scenarios across various driving amplitude levels ($\Omega_d$). This table demonstrates the impact of multiple refinement techniques on coherent error reduction, including standard RWA, RWA with full period adjustments, RWA with time-dependent corrections plus zero crossing, and RWA with various levels of effective corrections. The errors decrease significantly as more sophisticated corrections are applied, highlighting the enhanced control precision at lower driving amplitudes.
\end{table}

A standard fourth-order Runge-Kutta solver is used to evaluate the dynamics accurately, with a relative accuracy set to $5\cdot10^{-9}$. We observed, at the end of the pulse, an error in the length of the Bloch vector from unity, measured as $1 - \sqrt{r_x^2+r_y^2+r_z^2} \simeq 10^{-7}$. This deviation is due to numerical inaccuracies. All errors we discuss in the following are at least an order of magnitude larger than these numerical inaccuracies and stem not from a changed length of the Bloch vector. Instead its orientation deviates from the intended south pole by an angle $\gamma$. The coherence error, proportional to $\sin\gamma \approx \gamma$, and the population errors, $1-\cos\gamma \approx \gamma^2/2$, are expected to be roughly the square of the coherence error.

Incorporating corrections to the RWA and ensuring that the pulse length aligns with the driving period $\tau=\omega_{\rm LO,res}^{-1}$ for square pulses, we observed a reduction in error by approximately $\Omega_d/\omega_q \approx 10^{-2}$. Additionally, we achieve minor error reductions by slightly adjusting the oscillator frequency using a Bloch-Siegert shift calculated as $c_{\rm eff}\cdot 0.75\cdot (\Omega_d / \omega_{\rm LO})^2$, where $c_{\rm eff} \approx 0.995$. Doubling the pulse amplitude led to a fourfold decrease in the coherent error, indicating that our approach effectively mitigates all linear components of the coherent error.

We observed similar errors using Gaussian pulses when adjusting their amplitude to make the pulse width comparable to the duration of the square pulse by reducing the pulse amplitude by half. Adjusting the pulse duration to a multiple of the period within the RWA framework did not significantly alter the coherent error. Corrections to the RWA using a time-dependent resonance frequency also did not lead to error reductions. However, utilizing an effective constant average resonance frequency lowered the error, although it remained linearly dependent on the pulse amplitude. A significant decrease was observed when the resonance frequency was optimally shifted to $c_1^{({\rm eff})}= 0.2188$. Further adjustments to match the pulse length with the period maximized error reduction, making it quadratic relative to the amplitude. These optimal correction factors also proved effective for the shifted Gaussian pulse, confirming our approach's consistency.

\section{$Y_{\pi/2}$ Gate Implementation}\label{xpihalf}

\begin{figure}
\centering
\includegraphics[width=8.5cm]{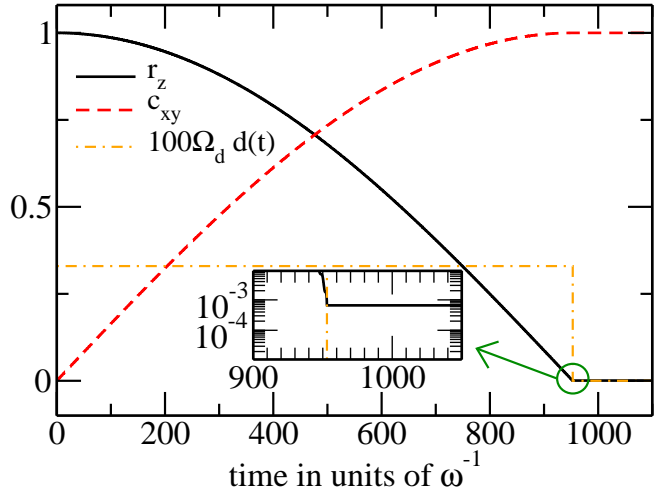}
\caption{\label{fig2} Numerical results showing the effect of a $\pi/2$-pulse implemented using a square pulse. The black full line in the main figure depicts the transition of the initial population from $r_z=1$ to $r_z=0$, indicating the erasure of the initial population. The red dashed line shows the resulting full coherence amplitude, $c_{xy}=1$, achieved at the pulse end (at time $953 \omega_q^{-1}$). The inset details the population error observed at the end of the pulse, demonstrating the deviations caused by the RWA.}
\end{figure}

Implementing a $\pi/2$-pulse involves adjusting the duration of square pulses or the width and duration of Gaussian pulses to achieve the desired control. This adjustment ensures that the time integral over the Rabi frequency sums to $\pi/2$. When a $\pi/2$-pulse acts on the ground state, the theoretical outcome should be zero population in $r_z$ and maximal coherence, $c_{xy}=1$, as depicted in Figure (\ref{fig2}). However, a coherent error typically manifests as a slight deviation of the Bloch vector from the equatorial plane, which makes the population error particularly sensitive to changes in pulse parameters.

\begin{table}[t]
    \centering
    \caption{$Y_{\pi/2}$-pulse Population $r_z$: \label{table:ypihalf}}
    \begin{tabular}{|d|b|b|b|} \hline
       \rowcolor{LightCyan}
       Square $\pi/2$ - Pulse & $\Omega_d=0.2\Omega_d^{(max)}$ & $\Omega_d=0.1\Omega_d^{(max)}$ & $\Omega_d=0.05\Omega_d^{(max)}$ \T\B \\ \hline
       RWA  &  $2.8\cdot 10^{-3}$ &  $1.5\cdot 10^{-3}$ & $6.6\cdot 10^{-4}$ \T\B \\ \hline
       RWA, full periods &  $2.6\cdot 10^{-5}$ &  $6.5\cdot 10^{-6}$ & $1.4\cdot 10^{-6}$ \T\B \\ \hline
       RWA + corrections &  $2.8\cdot 10^{-3}$ &  $1.5\cdot 10^{-3}$ & $6.6\cdot 10^{-4}$ \T\B \\ \hline
       RWA + corrections, full periods &  $7.2\cdot 10^{-5}$ &  $1.8\cdot 10^{-5}$ & $4.2\cdot 10^{-6}$ \T\B \\ \hline
       RWA + eff. corrections, full periods &  $4.1\cdot 10^{-5}$ &  $10^{-5}$ & $2.3\cdot 10^{-6}$ \T\B \\ \hline
       \rowcolor{LightCyan}
       Gaussian $\pi/2$ pulse & & &  \T\B \\ \hline
       RWA  & $7\cdot 10^{-5}$ & $6\cdot 10^{-5}$ & -- \T\B \\ \hline
       RWA, full periods & $9\cdot 10^{-6}$ & $2\cdot 10^{-6}$ & -- \T\B\\ \hline
       RWA + eff. corrections, full periods & $3.4\cdot 10^{-7}$ & $7\cdot 10^{-7}$ & -- \T\B \\ \hline
    \end{tabular}
    Assessment of population $r_z$ after applying a Square and Gaussian $\pi/2$-pulse at different normalized driving amplitudes ($\Omega_d$) using several approaches to enhance accuracy. This Table contrasts the errors obtained using the basic Rotating Wave Approximation (RWA), RWA with full period adjustments, and RWA supplemented by various corrections. Each row illustrates how different levels of refinement affect the accuracy of quantum state manipulation, particularly showing significant error reduction when full periods and effective corrections are applied.
\end{table}

Surprisingly, our experiments show that modifications to the RWA have minimal impact on reducing population errors. However, aligning the pulse duration to a multiple of the driving period can reduce errors by approximately two orders of magnitude, as detailed in Table \ref{table:ypihalf} for square $\pi/2$-pulses.

We observed a similar trend for Gaussian pulses, which is detailed in the same Table. Notably, significant error reductions are achieved simply by aligning the pulse duration to a multiple of the pulse period. The error typically scales as the square of the pulse amplitude. Nonetheless, we observed a further reduction in error when adjusting the resonance frequency to a value slightly red-shifted from the RWA expectation—opposite to the typical blue-shift from higher-order RWA corrections—specifically to $c_1^{({\rm eff})}= -0.084$. This adjustment shows that, while the error scales with pulse amplitude, it becomes indistinguishable from the numerical error under these optimized conditions, indicating a significant enhancement in precision.

\section{Implementation of initial state preparation}\label{sec:stateprep}

We explicitly tested an implementation of a pulse sequence $Y_{\pi/2} Z_{\pi - \theta} Y_{\pi/2}$ and varied $0\le\theta\le \pi$. The final coherence $c_{xy}$ and population $r_z$ are expected to hold
\begin{equation}
 c_{xy} = \sin\theta \quad\mbox{and}\quad r_z = \cos\theta .
\end{equation}
These results imply that the dominant error in the state preparation varies with the angle $\theta$. To assess this, we calculated the total error 
\begin{equation}\label{eq:delta}
\delta =\sqrt{(c_{xy}-\sin\theta)^2 + (r_z-\cos\theta)^2}, 
\end{equation}
which serves as a metric for the fidelity of the implemented state relative to the ideal theoretical state.

During the analysis, focusing on the optimal settings, we observed that the implementation using two square $\pi/2$ pulses typically results in errors on the order of $10^{-6}$. However, for $\theta=\pi$, where the sequence effectively becomes a $\pi$-pulse, more significant errors ranging from $10^{-5}$ to $10^{-4}$ were noted—this aligns with the errors previously observed for single $\pi$ pulses.

Interestingly, the effective correction factor, which adjusts for systematic errors in the pulse implementation, strongly depends on $\theta$. It varied from $0.996$ at $\theta = 0$ and $\theta = \pi$, to $0.332$ at $\theta = \pi/2$ (cf. Table \ref{table:state_preparation}). In contrast, sequences implemented with Gaussian pulses exhibited approximately twice the error magnitude of square pulses. Yet, the effective correction factor for Gaussian pulses was markedly different, averaging $0.115$ and showing little variation with $\theta$, indicating a near uniformity in the adjustment required across different angles within the accuracy of our measurements.

\begin{table}[t]
    \centering
    \caption{Total error $\delta$ for state preparation, Eq. \eqref{eq:delta}: \label{table:state_preparation}}
    \begin{tabular}{|d|b|b|b|b|} \hline
        \rowcolor{LightCyan}
        Pulse Shape & $\Omega_d=0.2\Omega_d^{(max)}$ & $\Omega_d=0.1\Omega_d^{(max)}$ & $\Omega_d=0.05\Omega_d^{(max)}$ \T\B \\ \hline
       Square   & $1-1.5\cdot 10^{-6}$ & $5-8\cdot 10^{-7}$ & $2-4\cdot 10^{-7}$ \T\B \\ \hline
       (shifted) Gaussian  & $2-4\cdot 10^{-6}$ & $1-2.5\cdot 10^{-6}$ & $6-10\cdot 10^{-7}$ \T\B \\ \hline
    \end{tabular}
    Total error $\delta$ for state preparation using square and shifted Gaussian pulses at various normalized driving amplitudes ($\Omega_d$). This table presents a range of errors for each pulse shape and driving strength, illustrating the error variability and the effectiveness of each pulse configuration in minimizing state preparation errors.
\end{table}

\section{Conclusions} \label{sec:conclusions}

In this work, we have explored the design and optimization of quantum pulses, a required element for advancing quantum computing technology. We aimed to enhance the fidelity and effectiveness of quantum gates and state preparation by implementing and analyzing various pulse configurations, including Square and Gaussian envelopes.

We have demonstrated improving basic pulse performance through design and adjustment. Our results revealed that coherent errors, which significantly degrade the performance of quantum operations, often stem from undesired terms in the Hamiltonian and several underlying assumptions in our experimental framework. By addressing these issues directly, we have identified the sources of errors and implemented strategies to mitigate them.

One key strategy was substantially reducing coherent errors by fine-tuning the external frequency and adjusting the pulse duration to a multiple of the inverse frequency. The later resulted in a stroboscopic type dynamics which severely optimizes the control over quantum states. This finding underscores the importance of precise frequency and duration control in pulse design.

Our analysis showed that even the most straightforward two-level system (TLS) models are not immune to coherent errors originating from naive design approaches. This insight is crucial for developing more accurate pulse schemes to mitigate the accumulation of coherent errors across large-scale quantum circuits.

Our results are agnostic to the quantum architecture, indicating the broad applicability of our findings. The techniques and improvements we have developed can be seamlessly integrated into diverse quantum computing platforms, including ion-trap, atomic, and photonic infrastructures.
The methodologies refined and validated in this study enhance the performance of quantum gates and provide a robust framework for the scalable implementation of quantum operations across various platforms.

\backmatter

\bmhead{Acknowledgements}
V.L.O was supported as part of the ASCR Quantum Testbed Pathfinder Program at Oak Ridge National Laboratory under FWP $\#$ ERKJ332. 

 \section*{Declarations}
 \subsection*{Conflict of Interest}
 The authors declare that they have no conflict of interest.
 \subsection*{Consent for publication}
 This manuscript has been authored by UT-Battelle, LLC, under Contract No. DE-AC0500OR22725 with the U.S. Department of Energy. The United States Government retains and the publisher, by accepting the article for publication, acknowledges that the United States Government retains a non-exclusive, paid-up, irrevocable, world-wide license to publish or reproduce the published form of this manuscript, or allow others to do so, for the United States Government purposes. The Department of Energy will provide public access to these results of federally sponsored research in accordance with the DOE Public Access Plan.
 \subsection*{Author Contribution}
 V.L.O. and P.N. developed the theoretical formalism. P.N. also developed and supervised the numerical experiments. A.W. and J.B. executed the simulations and collected the data. V.L.O. and P.N. analyzed and interpreted the numerical data and suggested additional experiments. All authors contributed to writing and revising the manuscript.

\begin{appendices}
\section{Effective Hamiltonian}
The Hamiltonian
\begin{equation}
    H = \left(-\frac{\omega_{q}}{2} \sigma_{z}+\Omega_{d}D(t)\sigma_{x}\right) 
\end{equation}
can be split in $H=H_0+H_1$ with
\begin{equation}
    H_0 = -\omega_{\rm LO} \frac{\sigma_z}{2} \quad\mbox{and}\quad 
    H_1 = - \delta \frac{\sigma_{z}}{2} +\Omega_{d}D(t)\sigma_{x}
\end{equation}
and $\delta = \omega_q - \omega_{\rm LO}$. 

The dynamics for the statistical operator $\rho(t)$ is described by the von-Neumann equation $\partial_t \rho(t) = i[\rho,H]$. 
Switching into a reference frame corotating with $H_0$, i.e. $\rho(t)=e^{-iH_0 t }\bar{\rho}(t) e^{iH_0 t}$ with $\bar{\rho}(t)$ the statistical operator in the rotating frame leads to a von-Neumann equation $\partial_t \bar{\rho}(t) = i[\bar{\rho}(t), \bar{H}]$ with
\begin{eqnarray}
   \bar{H} &=& e^{iH_0 t} H e^{-iH_0 t} - H_0 = e^{iH_0 t} H_1 e^{-iH_0 t} \nonumber \\
    &=& - \delta \frac{\sigma_{z}}{2} +\Omega_{d}d(t) \sin(2\omega_{\rm LO} t) \frac{\sigma_{x}}{2} + \Omega_{d}d(t) [1-\cos(2\omega_{\rm LO} t)] \frac{\sigma_{y}}{2} 
\end{eqnarray}

The Hamiltonian is periodic in time, i.e. $H(t+\tau)=H(t)$ and $\bar{H}(t+\tau)=\bar{H}(t)$, with $\tau=\omega_{\rm LO}^{-1}$. Thus, for the stroboscopic time evolution at multiples of $\tau$,an effective Hamiltonian must exist with
\begin{equation}
    e^{-iH_{\rm eff}N\tau} = {\cal T} e^{-i\int_0^{N\tau} dt \bar{H}(t)} .
\end{equation}
Since $H_{\rm eff}$ is independent of $N$ all terms on the right-hand side proportional to $N\tau$ form a series expansion of $H_{\rm eff}$, i.e.
\begin{equation}
    H_{\rm eff}^{(0)} = \frac{1}{N\tau} \int_0^{N\tau} dt \bar{H}(t) \quad \mbox{and} \quad 
    H_{\rm eff}^{(1)} = \frac{-i}{N\tau} \int_0^{N\tau} dt_1 \int_0^{t_1} dt_2 \bar{H}(t_1) \bar{H}(t_2)
\end{equation}
where only terms independent on $N\tau$ are included. Since integrals over multiples of periods of sine and cosine functions vanish, we find (the standard RWA result)
\begin{equation}\label{heff0}
    H_{\rm eff}^{(0)} = - \delta \frac{\sigma_{z}}{2} +\Omega_{d}d(t) \frac{\sigma_{y}}{2} .
\end{equation}
Only for $|d(t+\tau)-d(t)|\ll \Omega_d/\omega_{\rm LO}$, corrections due to the time dependence of $d(t)$ are smaller than contributions to $H_{\rm eff}^{(1)}$. \footnote{The pulse profile $d(t)$ in Eq.(\ref{heff0}) is, thus, a coarse-grained profile which reflects a kind of average within each single period. As long as its time-dependence is sufficient slow, the difference to $d(t)$ is minimal. Technically, the integral must be divided into parts over single periods in which $d(t)$ is constant. The result is then not determined by the parts proportional to $N\tau$ but to $\tau$.}

To determine the first corrections to the RWA result, i.e. $H_{\rm eff}^{(1)}$, we fix
\[ 
   h_1(t) = - \delta \frac{\sigma_{z}}{2} + \Omega_{d}d(t) \frac{\sigma_{y}}{2}  \; ,\; 
   h_2(t) = \Omega_{d}d(t) \sin(2\omega_{\rm LO} t) \frac{\sigma_{x}}{2} \; , \;
   h_3(t) = - \Omega_{d}d(t) \cos(2\omega_{\rm LO} t)  \frac{\sigma_{y}}{2} 
\]
and assuming very slow changing $d(t)$,thus, effectively time-independent $h_1$. With 
\[
A_{ij} = \int_0^{N\tau} dt_1 \int_0^{t_1} dt_2 h_i(t_1) h_j(t_2) 
\]
we find only the following contributions proportional to $N\tau$:
\begin{eqnarray}
  A_{21} &=& \Omega_{d}d(t) \frac{\sigma_{x}}{2} h_1\int_0^{N\tau} dt_1   \sin(2\omega_{\rm LO} t_1)  t_1  =  \Omega_{d}d(t) \frac{\sigma_{x}}{2} h_1 \cdot \frac{-N\tau}{2\omega_{\rm LO}}  \nonumber \\
  A_{12} &=& h_1 \Omega_{d}d(t) \frac{\sigma_{x}}{2} \int_0^{N\tau} dt_1   \frac{1-\cos(2\omega_{\rm LO} t_1)}{2\omega_{\rm LO}}  =  h_1 \Omega_{d}d(t) \frac{\sigma_{x}}{2}  \cdot \frac{N\tau}{2\omega_{\rm LO}}  \nonumber \\
  A_{23} &=& -[\Omega_{d}d(t)]^2 \frac{\sigma_x \sigma_{y}}{4} \int_0^{N\tau} dt_1   \sin(2\omega_{\rm LO} t_1)  \int_0^{t_1} dt_2  \cos(2\omega_{\rm LO} t_2) \nonumber \\
  &=& -[\Omega_{d}d(t)]^2 \frac{\sigma_x \sigma_{y}}{4} \int_0^{N\tau} dt_1   \frac{\sin^2(2\omega_{\rm LO} t_1)}{2\omega_{\rm LO}}   
  = -[\Omega_{d}d(t)]^2 \frac{\sigma_x \sigma_{y} N\tau }{16\omega_{\rm LO}}  \nonumber \\
 A_{32} &=& -[\Omega_{d}d(t)]^2 \frac{\sigma_y \sigma_{x}}{4} \int_0^{N\tau} dt_1   \cos(2\omega_{\rm LO} t_1)  \int_0^{t_1} dt_2  \sin(2\omega_{\rm LO} t_2) \nonumber \\
  &=& -[\Omega_{d}d(t)]^2 \frac{\sigma_y \sigma_{x}}{4} \int_0^{N\tau} dt_1  \cos(2\omega_{\rm LO} t_1) \frac{1- \cos(2\omega_{\rm LO} t_1)}{2\omega_{\rm LO}}   
  = [\Omega_{d}d(t)]^2 \frac{\sigma_y \sigma_{x} N\tau }{16\omega_{\rm LO}}  \nonumber 
\end{eqnarray}
leading to
\begin{equation}
 H_{\rm eff}^{(1)} =  - \frac{3 (\Omega_d d(t))^2}{4 \omega_{\rm LO}} \frac{\sigma_x}{2}
- \Omega_d  d(t)  \frac{\delta}{2\omega_{\rm LO}} \frac{\sigma_y}{2} .
\end{equation}

\end{appendices}

\bibliography{refs}

\end{document}